\documentclass[twocolumn,showpacs,floatfix,nofootinbib]{revtex4-1}
\usepackage{graphicx}
\usepackage{bm} 
\usepackage{amssymb} 
\usepackage{amsmath} 
\usepackage{braket} 

\usepackage{natbib} 
\usepackage{hyperref}

\begin{document}
\title{Effects of a tilted magnetic field in a Dirac double layer}

\author{Sergey S. Pershoguba$^{1,2}$}
\author{D. S. L. Abergel$^1$}
\author{Victor M. Yakovenko$^2$}
\author{A. V. Balatsky$^{1,3}$}

\affiliation{$^1$Nordita, KTH Royal Institute of Technology and Stockholm University, Roslagstullsbacken 23, SE-106 91 Stockholm, Sweden}
\affiliation{$^2$Condensed Matter Theory Center, Department of Physics, University of Maryland, College Park, Maryland 20742-4111, USA}
\affiliation{$^3$Institute for Materials Science, Los Alamos National Laboratory, Los Alamos, New Mexico 87545, USA}

\date{\today} 

\begin{abstract}
We calculate the energy spectrum of a Dirac double layer, where each layer has the Dirac electronic dispersion, in the presence of a tilted magnetic field and small interlayer tunneling.  We show that the energy splitting between the Landau levels has an oscillatory dependence on the in-plane magnetic field and vanishes at a series of special tilt angles of the magnetic field.  Using a semiclassical analysis, we show that these special tilt angles are determined by the Berry phase of the Dirac Hamiltonian.  The interlayer tunneling conductance also exhibits an oscillatory dependence on the magnetic field tilt angle, known as the angular magnetoresistance oscillations (AMRO).  Our results are applicable to graphene double layers and thin films of topological insulators.
\end{abstract}


\maketitle

\section{Introduction} \label{sec:Intro}

Recently there has been considerable interest in the effects of a magnetic field in materials with the Dirac dispersion in the electronic energy spectrum\cite{Wehling2014}.  Most studies focus on a perpendicular magnetic field applied to a two-dimensional (2D) Dirac material, e.g.,\ graphene\cite{Abergel2010} or the surface of a topological insulator (TI).  The Landau quantization of Dirac fermions produces the unconventional quantum Hall effect \cite{Gusynin2005}, which is often taken as an experimental signature of Dirac fermions in the system \cite{Kim-QHE,Novoselov-QHE}. In addition, a number of papers consider the case with an in-plane component of the magnetic field\cite{Pershoguba2010,Zyuzin2011,Pershoguba2012,Pratley2013,Zhang2012,Lin2013,Latyshev2013,Mishchenko2014,He2014,Choi2011,*Hyun2012,Goncharuk2012a,*Goncharuk2012,Taskin2010,*Taskin2011}.  The in-plane component produces a relative shift in momentum space of the Dirac cones in adjacent layers.  This effect results in an unusual energy spectrum and dependence of the interlayer tunneling current on the magnetic field \cite{Pershoguba2010,Latyshev2013,Zyuzin2011,Pershoguba2012,Pratley2013}. Magnetoresistance and tunneling spectroscopy for the in-plane magnetic field were measured in thin films of TIs \cite{Zhang2012,Lin2013}, a graphite mesa \cite{Latyshev2013}, and a graphene double layer \cite{Mishchenko2014}.  A relative twist of the layers in a graphene bilayer also produces an effect similar to the in-plane magnetic field \cite{CastroNeto,Andrei,He2014}.  The Landau levels in a tilted magnetic field were studied for graphene multilayers \cite{Choi2011,*Hyun2012,Goncharuk2012a,*Goncharuk2012}. An unusual dependence of the resistance on the magnetic field orientation was found in a bulk TI \cite{Taskin2010,*Taskin2011}.

\begin{figure} \centering
(a) \includegraphics[width=0.9\linewidth]{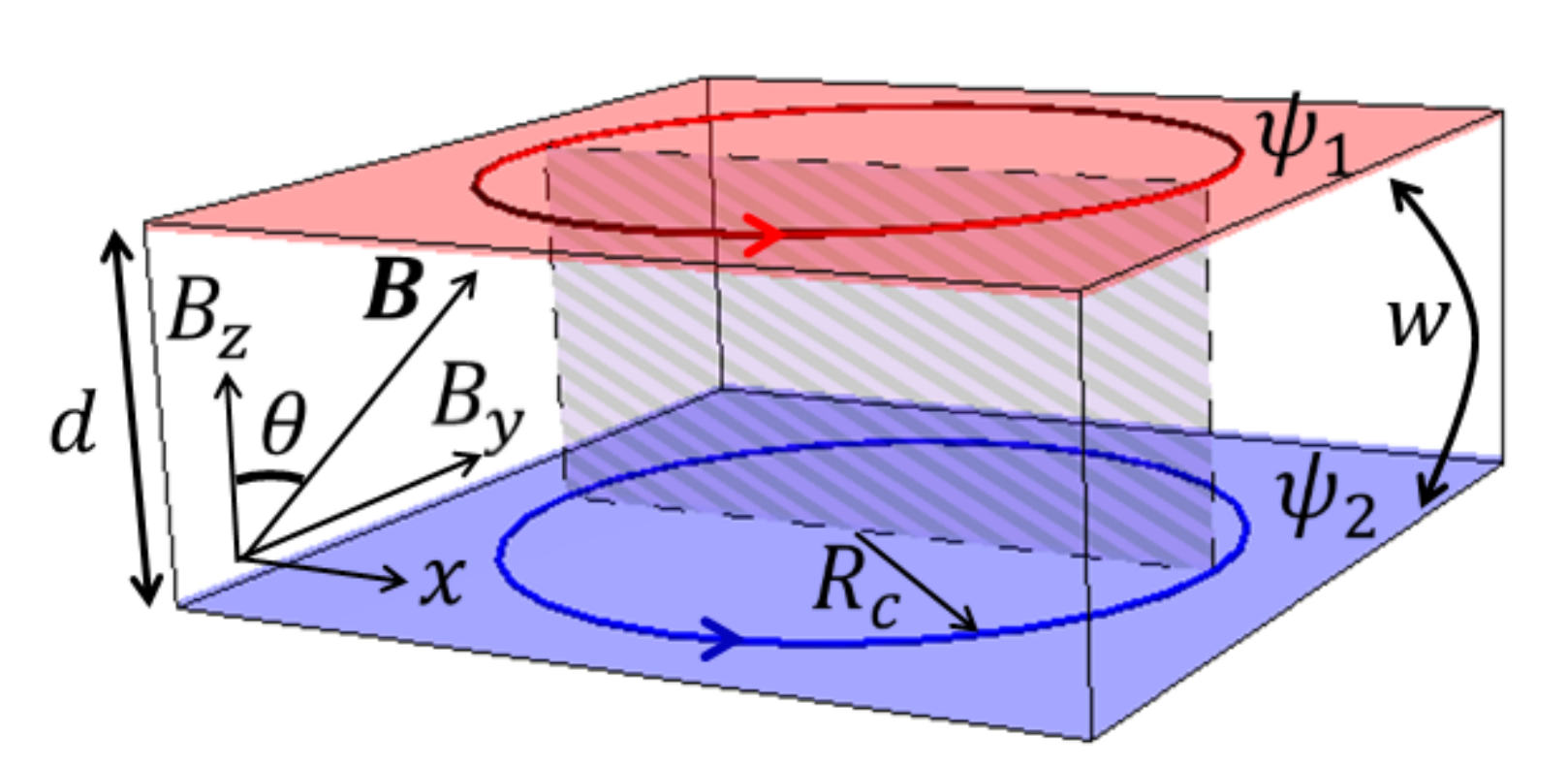} \\
(b) \includegraphics[width=0.9\linewidth]{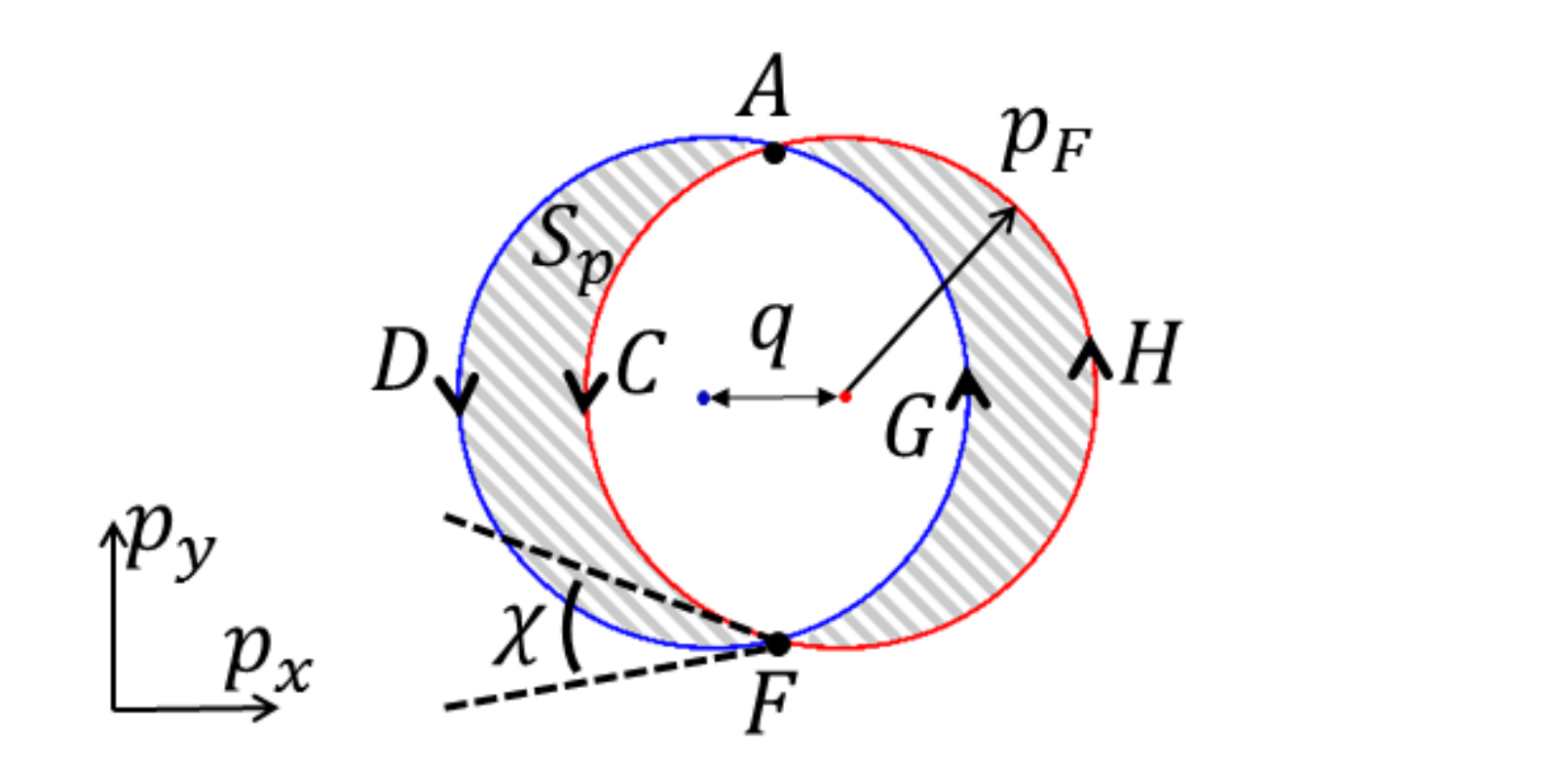}
\caption{(color online) (a) Double layer of thickness $d$ in the tilted magnetic field $\bm B = (0,B_y,B_z)$.  The out-of-plane magnetic field $B_z$ induces the in-plane cyclotron motion of the radius $R_c$. Interference between the two orbits is controlled by the flux of the in-plane magnetic field $B_y$ through the Aharonov-Bohm area shown as the shaded rectangle. (b) Semiclassical electron orbits in momentum space in the two layers are shifted by $q=eB_yd$. Interference between the orbits is controlled by the shaded areas $S_p$. Both real (a) and momentum (b) space pictures show that the interlayer tunneling $w$ is suppressed at the magic angles $\theta_N$ in Eq.~(\ref{classic}). } \label{fig:Bilayer}
\end{figure}

The oscillatory dependence of resistance on the orientation of a tilted magnetic field, called the angular magnetoresistance oscillations (AMRO), was first observed in organic conductors \cite{Kartsovnik1988}.  AMRO are characteristic for layered materials, such as organic conductors \cite{Singleton}, intercalated graphite \cite{Iye1994,Enomoto2006}, Sr$_2$RuO$_4$ \cite{AMRO-Sr2RuO4,AMRO-Sr2RuO4b,Bergemann2003}, and high-$T_c$ cuprates \cite{AMRO-cuprates-Yakovenko,AMRO-cuprates-Hussey,Analytis2007} (see more references in Refs.~[\onlinecite{McKenzie,*Moses1999,Yakovenko2006,Cooper2006}]).  AMRO are manifested as resistivity oscillations periodic in $\tan\theta=B_y/B_z$, where the tilt angle $\theta$ is expressed in terms of the in-plane $B_y$ and out-of-plane $B_z$ components of the magnetic field.  The effect is distinct from the usual quantum oscillations, which are periodic in $1/B_z=1/B\cos \theta$.   Although AMRO were originally studied for an infinite layered crystal \cite{AMRO-Yamaji}, it was later shown that the effect {exists} even for two layers \cite{McKenzie,*Moses1999}.  AMRO can be interpreted in terms of the interlayer Aharonov-Bohm (AB) effect in the following way\cite{Cooper2006,Yakovenko2006}.  Consider a double layer of the distance  $d$ between the layers in the tilted magnetic field $\bm B = (0,B_y,B_z)$, as shown in Fig.~\ref{fig:Bilayer}(a).  The perpendicular magnetic field $B_z$ induces cyclotron motion of the radius $R_c=p_F/eB_z$ in each layer, where $p_F$ is the Fermi momentum, and $e$ is the electron charge.  The cyclotron diameter $2p_F$ and the interlayer distance $d$ form the area $S_{\rm AB}=2R_cd$ shown by the shaded rectangle in Fig.~\ref{fig:Bilayer}(a).  The flux of the in-plane magnetic field $B_y$ through this area determines the interference condition  $B_yS_{\rm AB}=2\pi\hbar(N+{\rm const})/e$ between electron trajectories involving in-plane cyclotron motion and interlayer tunneling.  For materials with the parabolic electronic energy spectrum, destructive interference suppresses interlayer tunneling \cite{Yakovenko2006} at the following ``magic'' angles $\theta_N$
\begin{equation}
  p_F d \tan\theta_N =\hbar\left(\pi N-\frac{\pi}{4}\right),\quad N = 1,2,\ldots .
\label{classic}
\end{equation}
Alternatively, the same condition can be obtained in the momentum space $\bm p=(p_x,p_y)$. The in-plane magnetic field $B_y$ shifts the relative momenta of the Fermi circles in the adjacent layers by \cite{Pershoguba2010,Pershoguba2012}
\begin{equation}
	\Delta \bm p_x = \bm q = \hat{\bm x}eB_yd \label{q},
\end{equation}
as illustrated in Fig.~\ref{fig:Bilayer}(b).  The perpendicular magnetic field $B_z$ induces cyclotron motion indicated by the arrows, and interference between the two circular orbits is controlled by the shaded areas $S_p\approx 2p_Fq$ shown in Fig.~\ref{fig:Bilayer}(b). The Onsager-like interference condition $S_p/eB_z = 2\pi\hbar(N+{\rm const})$ gives Eq.~(\ref{classic}) as well.

Although many Dirac materials have a layered structure, the effect of AMRO received limited attention for these materials. AMRO were measured in intercalated graphite compounds \cite{Iye1994,Enomoto2006}, and offsets $-0.39\pi$ and $-\pi/4$ in Eq.~(\ref{classic}) were observed. Recently, the effect of the Berry curvature, which may be present in gapped Dirac materials, on quantum oscillations and AMRO was studied in Refs.~[\onlinecite{Wright2013a},\onlinecite{Wright2013}].

Here we present a theoretical study of AMRO in the simplest case of the Dirac double layer, where each layer has a linear electronic energy spectrum.  It is realized experimentally for a double layer of graphene \cite{Mishchenko2014} or the opposite surfaces of a thin film of a TI\cite{Armitage,Zyuzin2011,Pershoguba2012}. In the presence of a small interlayer tunneling, we find that the Landau levels spectrum in a tilted magnetic field has angular dependence similar to AMRO.  The levels become doubly degenerate at the ``magic'' tilt angles $\theta_N$, where the effective interlayer coupling is suppressed due to the destructive AB interference. We also calculate the interlayer conductance, which exhibits both the Shubnikov-de Haas and AMRO oscillations.  We find a deviation from the standard $-\pi/4$ offset angle in Eq.~(\ref{classic}) and explain it semiclassically using the Berry phase.

\section{Hamiltonian of a double layer}

Consider a Dirac double layer of thickness $d$ as shown in Fig.~\ref{fig:Bilayer}(a). The Hamiltonian of the model in the second-quantized form is
\begin{eqnarray}
  && H_0 = \int d^2 p \left[{\psi_{\bm p}^1}^\dag h(\bm p) \psi_{\bm p}^1
  + \alpha\,{\psi_{\bm p}^2}^\dag h(\bm p) \psi_{\bm p}^2\right], \label{Ham} \\
  && h(\bm p) = v(\bm\sigma\cdot\bm p) = v(\sigma_xp_x+\sigma_yp_y).
\label{h}
\end{eqnarray}
Here, $\psi_{\bm p}^j$ is the wavefunction of an electron with in-plane momentum $\bm p=(p_x,p_y)$ on the opposite layers labeled by $j=1,2$, and $h(\bm p)$ is the Dirac Hamiltonian. We consider the simplest case where each layer contains only a single flavor of the Dirac electrons. However our analysis can be extended to multiple Dirac flavors per layer as, for example,\ in graphene, where the two flavors correspond to the valley and spin degree of freedom\cite{Abergel2010}.  The Pauli matrices $\bm \sigma$ act on the spinor wave functions $\psi^j=[\psi^j_{\uparrow},\psi^j_{\downarrow}]$, where the pseudospin index $\uparrow\downarrow$ corresponds to a sublattice degree of freedom in graphene and to the real spin in TIs.

The Hamiltonian $h(\bm p)$ has the Dirac cone linear energy dispersion $E_{\bm p} = \pm v |\bm p|$. The eigenstates corresponding to the positive and negative energies are the spinors
\begin{equation}
  \psi_{+,\bm p} = \frac{1}{\sqrt{2}}
  \left[\begin{array}{c} e^{-i\gamma} \\ 1 \end{array}\right],\quad
  \psi_{-,\bm p} = \frac{1}{\sqrt{2}}
  \left[\begin{array}{c} -1 \\ e^{i\gamma} \end{array}\right],
\label{wv}
\end{equation}
where $\gamma={\rm arctan}(p_y/p_x)$ is the angle of $\bm p$ in the 2D momentum space. The eigenstates (\ref{wv}) have parallel and antiparallel locking of the chiral pseudospin and the momentum, respectively. One can define the Berry phase for the wave functions in Eq.~(\ref{wv}). The winding of the Berry phase along an arbitrary contour $\mathcal C$ in the momentum space is
\begin{equation}
	\Gamma(\mathcal C) = i\int_{\mathcal C} d\bm p\,
  \langle \psi_{\pm,\bm p} \mid \partial_{\bm p} \mid \psi_{\pm,\bm p}\rangle
  = \pm\frac{\Delta \gamma}{2},
\label{berry}
\end{equation}
where $\Delta\gamma$ is the angle traced by $\mathcal C$ when viewed from the origin. Note that the wave functions in Eq.~(\ref{wv}) corresponding to positive and negative energies have opposite Berry phases. In Sec.~\ref{sec:Semiclassical}, we show that the Berry phase can change the magic angles offset in Eq.~(\ref{classic}).

In Eq.~(\ref{Ham}), the Dirac cones on the opposite layers have either the same $\alpha = 1$ or opposite $\alpha = -1$ chiralities. The case $\alpha = 1$ corresponds to a graphene double layer \cite{Mishchenko2014}, where the alignment of graphene lattices in the real space translates into the alignment of the Dirac cones of the same chirality in the momentum space. The case $\alpha = -1$ corresponds to a TI film \cite{Pershoguba2012}, where the Rashba vectors normal to the opposite surfaces of the film define the Dirac cones of the opposite chirality \cite{Pershoguba2012a}.

\section{Effect of a magnetic field } \label{sec:magfields}

Now let us introduce a perpendicular magnetic field $B_z$. With the Peierls substitution, the Dirac Hamiltonian becomes $h(\bm p-e\bm A)$, where we choose the Landau gauge $\bm A=-yB_z\hat{\bm x}$ for the vector potential $\bm A$.  The energy spectrum is given by the Landau levels labeled by the integer $n=0,\pm1,\ldots$
\begin{eqnarray}
  \Phi_{n,p_x} &=& \frac{1}{\sqrt{2}}
  \left[\begin{array}{c} \phi_{|n|,p_x}\\{\rm sgn}(n)\phi_{|n-1|,p_x}\end{array}\right],
  \quad E_n = {\rm sgn}(n)\frac{\hbar v\sqrt{2n}}{l}. \nonumber \\
  l &=& \sqrt{\hbar/eB_z}.
\label{eigenfunc1}
\end{eqnarray}
Here $l$ is the magnetic length, and $\phi_{m,p_x}$ are the usual harmonic-oscillator wave functions
\begin{equation*}
  \phi_{m,p_x}(y)=\frac{e^{-(y+p_xl^2/\hbar)^2/2l^2}}{\sqrt{2^m m! l\sqrt{\pi}}}
  H_m\left(\frac{y+p_xl^2/\hbar}{l}\right),
\end{equation*}
where $H_m$ are the Hermite polynomials. The momentum $p_x$ is a good quantum number and controls the position $y_c = -p_x/eB_z$ along the $\hat {\bm y}$ axis around which the wave functions $\phi_{m,p_x}$ are localized.

\begin{figure}  \centering
(a) \includegraphics[width=0.9\linewidth]{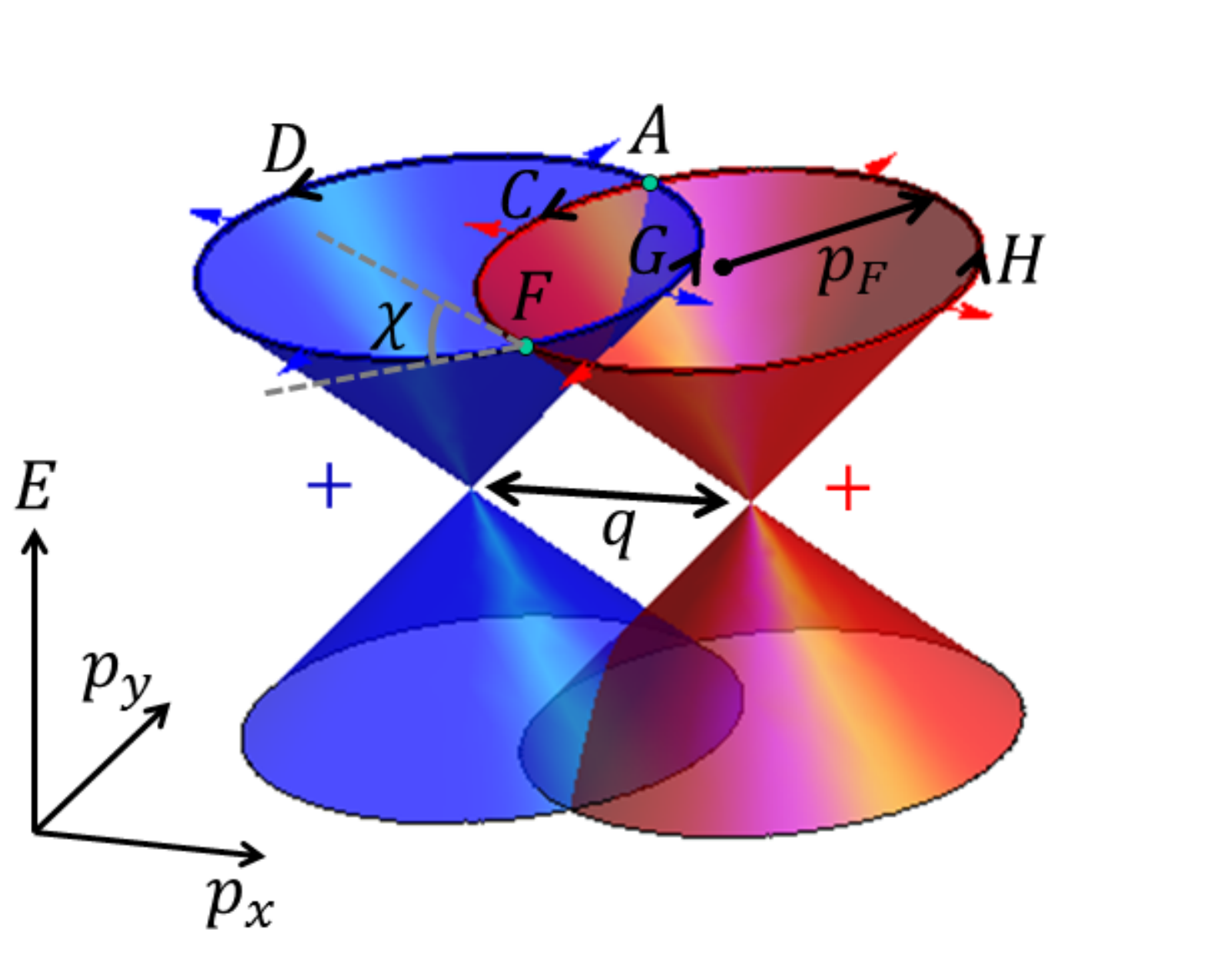}
\caption{(color online) Dirac cones of the two layers shifted in momentum space by $q=eB_yd$. The out-of-plane magnetic field $B_z$ induces cyclotron motion in the direction shown by the black arrows.  The two cyclotron orbits intersect at the angle $\chi$ at the points $A$ and $F$.  The red and blue arrows attached to the Fermi circles show the pseudospin direction for each Dirac cone for $\alpha = 1$ in Eq.~(\ref{Ham}).  Either blue or red arrows are reversed for $\alpha=-1$.} \label{fig:Cones}
\end{figure}

Next, let us turn on a parallel magnetic field $B_y$, so that the vector potential becomes $\bm A=(zB_y-yB_z)\,\hat{\bm x}$. For a single layer, the in-plane magnetic field does not have any orbital effect. But for a double layer, the term $-zB_y$ produces a relative shift of the in-plane momenta $\Delta p_x = q$ on the opposite layers \cite{Pershoguba2010,Pershoguba2012} given by Eq.~(\ref{q}).  The dynamics of electrons can be understood semiclassically as the cyclotron motion on the shifted Dirac cones corresponding to the opposite layers, as shown in Fig.~\ref{fig:Cones}. In the quantum description, the momentum $p_x$ controls the $y_c$ position around which the wave functions in Eq.~(\ref{eigenfunc1}) are localized. So, the shift $q$ in the momentum space also produces a relative shift of the wave functions in real space
\begin{equation}
  \Delta y= \frac{q}{eB_z} = d\,\frac{B_y}{B_z} = d\tan\theta.
\label{Dy}
\end{equation}
For simplicity, we do not include the Zeeman coupling of the magnetic field to the electron spins and leave it for future studies\footnote{The Zeeman energy $E_Z=g\mu_BB$ is linear in the magnetic field, whereas the energies of the Landau levels~(\ref{eigenfunc1}) scale as the square root $\sqrt B_z$.  Thus, the effect of the Zeeman coupling can be neglected for small enough magnetic field $B$.  For larger magnetic fields, the effect becomes noticeable and is different for graphene and TIs.  For graphene, the Zeeman coupling simply splits the Landau levels.  For TIs, the in-plane magnetic field $B_y$ generates a term $B_y\sigma_y$ in the Hamiltonian and, thus, shifts the Dirac dispersion in the 2D momentum space \cite{Pershoguba2012}.  On the other hand, the perpendicular magnetic field $B_z$ produces a term $B_z\sigma_z$ in the Dirac Hamiltonians~(\ref{h}) and, therefore, generates a gap.  A careful consideration of the Zeeman contribution can be done within our approach, but it complicates the discussion, so we leave it for future studies.}.

\section{Interlayer tunneling} \label{sec:coupling}

The spectrum of the Hamiltonian in Eq.~(\ref{Ham}) in the presence of the titled magnetic field consists of the Landau levels, which are double degenerate because of the identical Dirac Hamiltonians in the two layers. Now suppose the layers are coupled by the tunneling Hamiltonian
\begin{eqnarray}
  H_w = \int d^2p \left[{\psi_{\bm p}^1}^\dag\, W^\dag\, \psi_{\bm p}^2
  +{\psi_{\bm p}^2}^\dag\, W\,\psi_{\bm p}^1 \right],\,\, W = w\, \bm I.
\label{w}
\end{eqnarray}
In general, $W$ is the interlayer tunneling matrix in the pseudospin space \cite{Pershoguba2010}, but here we consider the simplest case where it is proportional to the unit matrix ${\bm I={\rm diag}(1,1)}$.  We also assume that the interlayer tunneling is local in real space, so the in-plane momentum $\bm p$ is conserved, and the amplitude $w$ does not depend on $\bm p$.

We expand the wave functions $\psi^1=\sum_n \psi^1_n \Phi_{n,p_x}$ and $\psi^2 = \sum_n \psi^2_n \Phi_{n,p_x-q}$ in the basis of the Landau functions~(\ref{eigenfunc1}), where the eigenvalue equation for the Hamiltonian $H_0+H_w$ in the tilted magnetic field becomes
\begin{align}
  \sum_m \left[\begin{array}{cc}
  (E_n-E)\delta_{nm} & w_{n,m} \\ w_{m,n} & (\alpha E_n-E)\delta_{nm}
  \end{array}\right]
  \left[\begin{array}{c} \psi_m^1 \\ \psi_m^2 \end{array}\right]=0.
\label{Schrod}
\end{align}
The matrix elements $w_{nm}= w\langle\Phi_{n,p_x}\mid\Phi_{m,p_x-q}\rangle$ of $H_w$ between the Landau functions on the opposite layers are
\begin{align}
  \frac{w_{n,m}}{w}
  = & -\frac{e^{-\beta^2/2} (-\beta)^{|n|-|m|}}{2^\eta} \left[\sqrt{\frac{|m|!}{|n|!}}
  L^{(|n|-|m|)}_{|m|}\left(\beta^2\right)\right.
\nonumber\\
  & +{\rm sgn}(nm)  \left. \sqrt{\frac{(|m|-1)!}{(|n|-1)!}} L^{(|n|-|m|)}_{|m|-1}
  \left(\beta^2\right)\right],
\label{wnm} \\
  \beta= & \frac{q l}{\hbar\sqrt{2}}= B_yd\sqrt{\frac{e}{2\hbar B_z}}.
\label{beta}
\end{align}
Here $L_j^{(k)}(x)$ are the Laguerre polynomials, and the exponent is $\eta=0$, 1/2, and 1 for the cases $n=m=0$, $|n|>m=0$, and $|n|\ge |m|>0$, respectively.  The matrix elements (\ref{wnm}) are derived in Appendix \ref{Sec-MatrixElements}. Note that the two-component spinor structure of the wave functions~(\ref{eigenfunc1}) produces the two terms with the Laguerre functions in Eq.~(\ref{wnm}). In the case of a simple parabolic spectrum, the analogous matrix elements have only one such term \cite{MacDonald1992}.

\section{Discussion of the spectrum}  \label{sec:spectrum}

\begin{figure}
\centering
(a) \includegraphics[width=.85\linewidth]{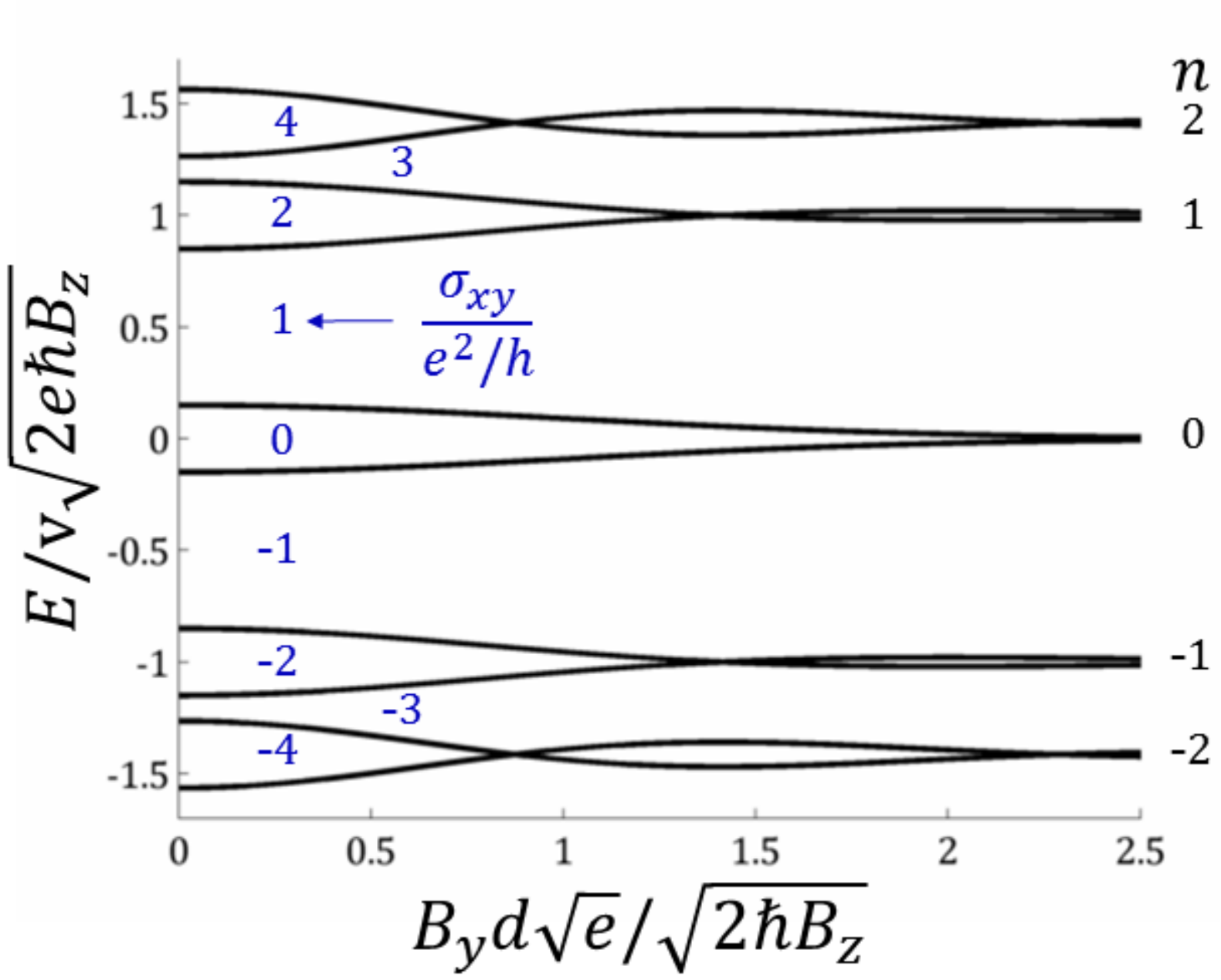} \\
(b) \includegraphics[width=.85\linewidth]{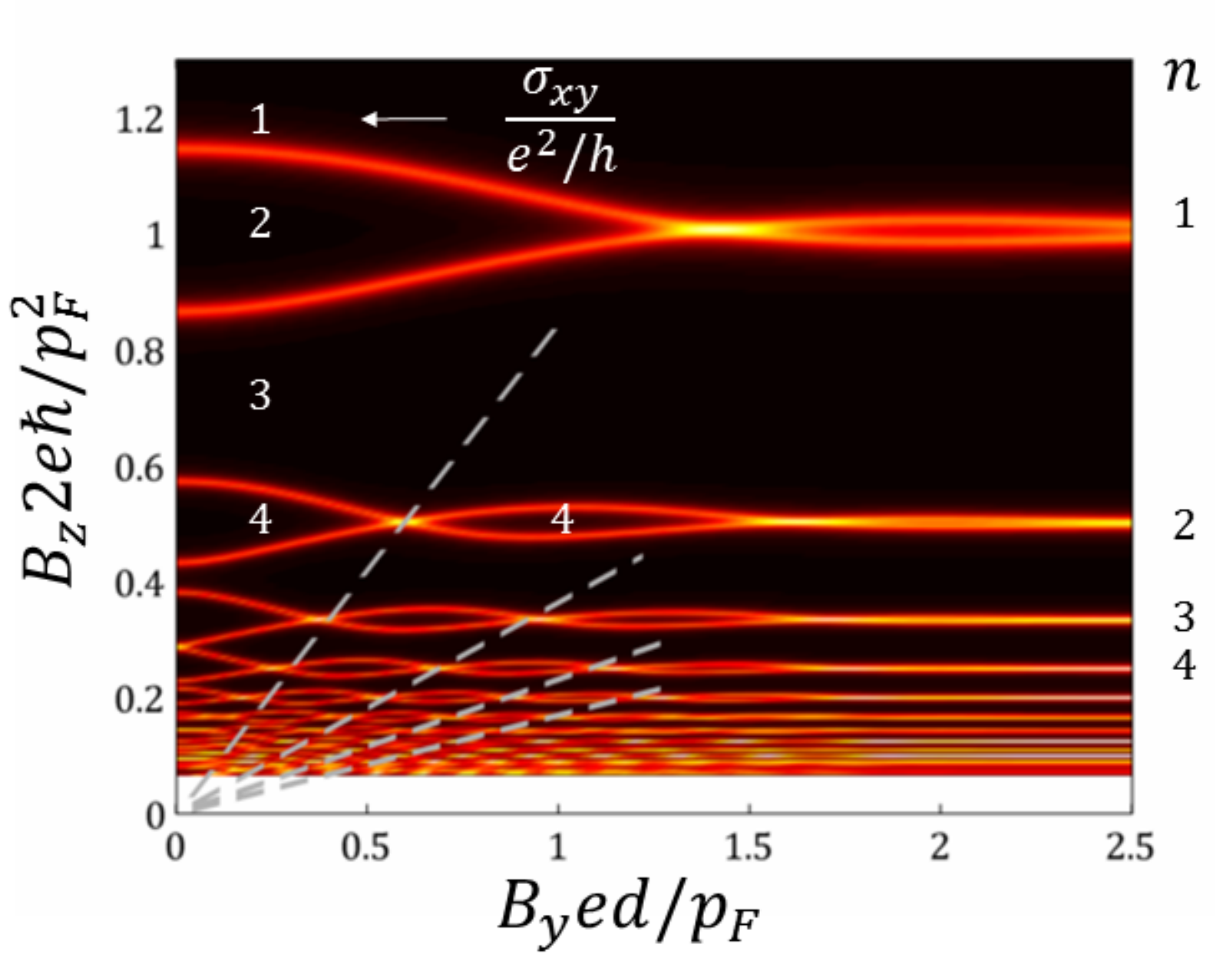} \\
(c)  \includegraphics[width=.85\linewidth]{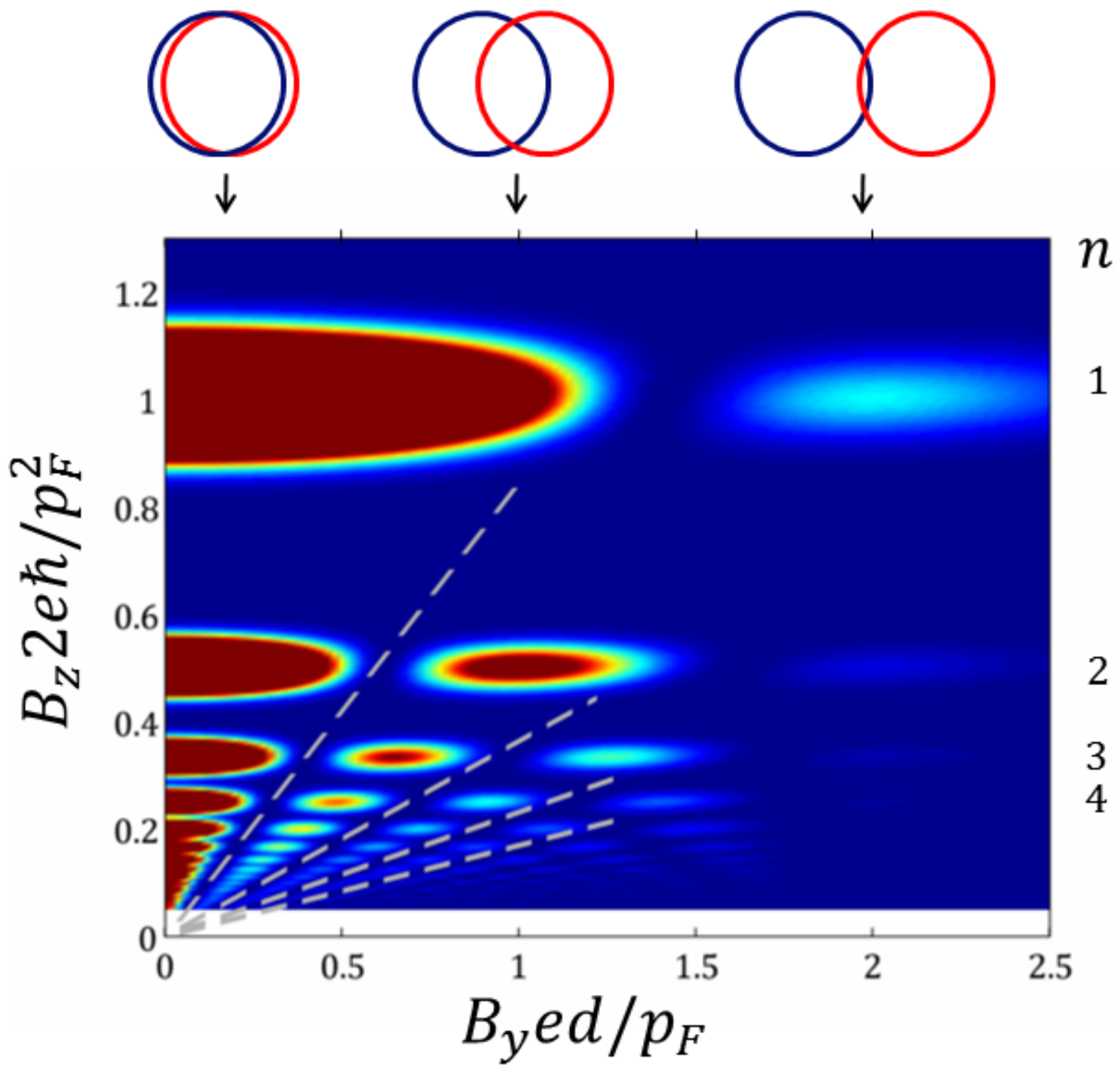}
\caption{(color online) (a) The energy spectrum of Eq.~(\ref{Schrod}) for $\alpha=1$ vs.\ $B_y$ for a fixed $B_z$. The number on the right axis is the Landau level index $n$.  The numbers on the plot indicate the filling factor $\nu$, which defines the quantum Hall conductivity. (b) Density of states (DOS) at the Fermi energy $E_F=vp_F$ plotted vs.\ $B_y$ and $B_z$.  Dashed lines correspond to the magic tilt angles given by Eq.~(\ref{zeros1}).  (c) The out-of-plane conductance $G_{zz}$ from Eq.~(\ref{tunnel}) vs.\ $B_y$ and $B_z$.  The Fermi circles of the two layers shifted by $q=B_yed$ are shown at the top.}
\label{fig:same}
\end{figure}

\begin{figure}
\centering
(a) \includegraphics[width=0.85\linewidth]{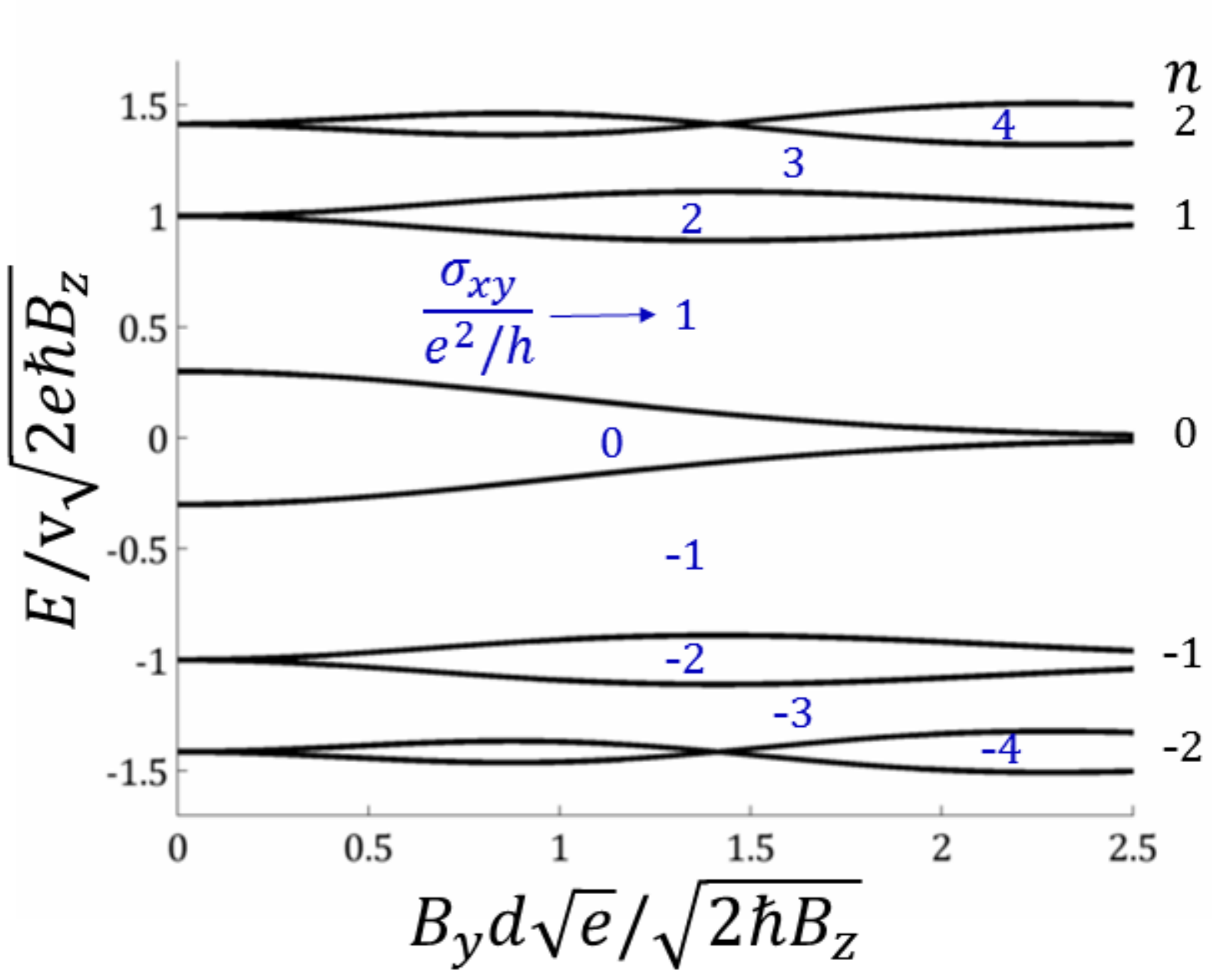} \\
(b) \includegraphics[width=0.85\linewidth]{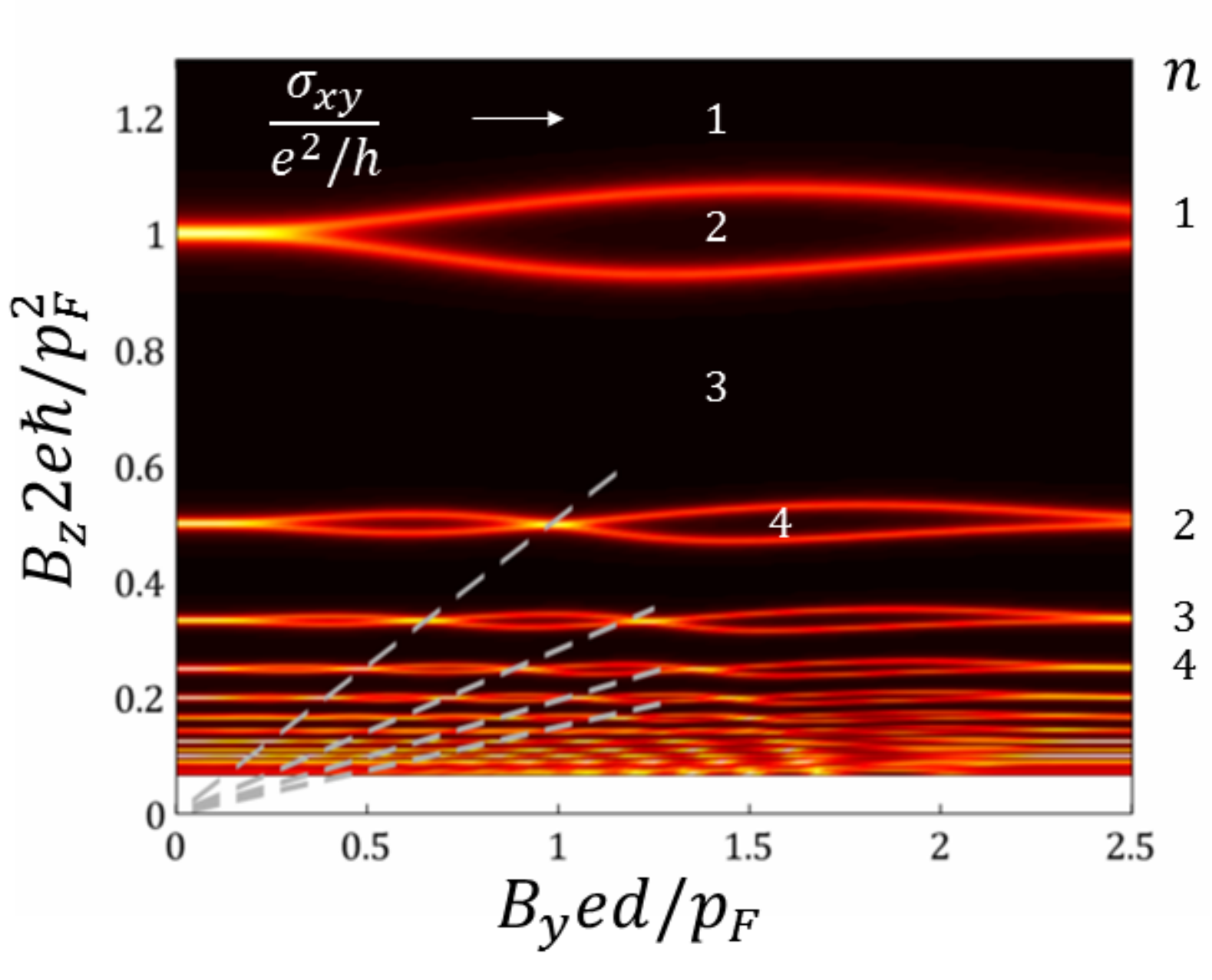}  \\
(c) \includegraphics[width=0.85\linewidth]{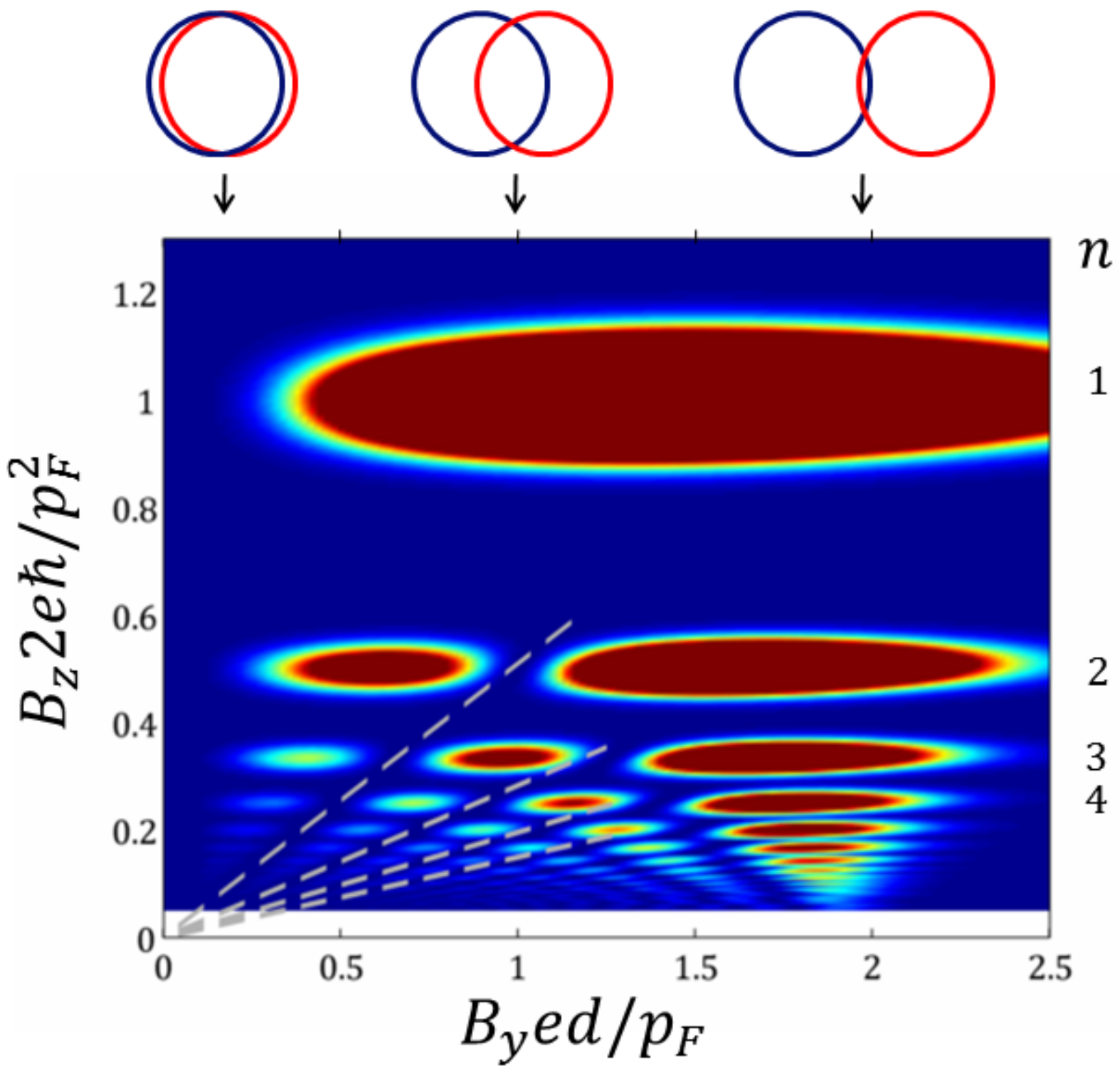}
\caption{(color online) (a) The energy spectrum of Eq.~(\ref{Schrod}) for $\alpha=-1$ vs.\ $B_y$ for a fixed $B_z$. The number on the right axis is the Landau level index $n$.  The numbers on the plot indicate the filling factor $\nu$, which defines the quantum Hall conductivity. (b) DOS at the Fermi energy $E_F=vp_F$ plotted vs.\ $B_y$ and $B_z$.  Dashed lines correspond to the magic tilt angles given by Eq.~(\ref{zeros2}).  (c) The out-of-plane conductance $G_{zz}$ from Eq.~(\ref{tunnel}) vs.\ $B_y$ and $B_z$.  The Fermi circles of the two layers shifted by $q=B_yed$ are shown at the top.} \label{fig:opposite}
\end{figure}

We calculate the energy spectrum in a tilted magnetic field by solving Eq.~(\ref{Schrod}) numerically and show the results\footnote{The interlayer tunneling amplitude $w$ is $0.15v\sqrt{2e\hbar B_z}$ in Fig.~3(a), $0.3v\sqrt{2e\hbar B_z}$ in Fig.~4(a), and $0.05E_F$ for panels (b) and (c) in Figs.~3 and 4.  We assume that DOS of the Landau levels has finite width $\Gamma$ as defined in Eq.~(\ref{lldos}). We use $\Gamma=0.1w$ for panels (b) and $\Gamma = w$ for panels (c).  In order to enhance contrast in panels (c), we clip the color map at $5\%$ of its maximal value. Namely, we plot $G_{zz}(B_y,B_z)$ for $G_{zz}(B_y,B_z)<M$ and $M$ for $G_{zz}(B_y,B_z)>M$, where $M=0.05\, {\rm max}\left[ G_{zz}(B_y,B_z)\right]$.} for the same $\alpha=1$ and opposite $\alpha=-1$ chiralities in Figs.~\ref{fig:same} and \ref{fig:opposite}.  Panels (a) show the energy levels $E$ versus the in-plane magnetic field $B_y$ for a fixed $B_z$.  The Landau level index $n$ is shown on the right vertical axis.  We observe splitting of the Landau levels, which oscillates as a function of $B_y$.  This behavior can be understood using perturbation theory in $w$.  For $w=0$, the wave functions $\Phi_{n}$ and $\Phi_{\alpha n}$ localized on different layers have the same energy $E_n$ according to Eq.~(\ref{eigenfunc1}).  To the first order in $w$, the symmetric-antisymmetric (SAS) splitting of the Landau levels is given by the matrix elements $w_{n,\alpha n}$:
\begin{align}
  E_n^\pm = E_n \pm w_{n,\alpha n}, \;\;
  w_{n,\alpha n} =  w\langle\Phi_{n,p_x}\mid\Phi_{\alpha n,p_x-q}\rangle.
\label{sas}
\end{align}
The wave functions $\Phi_{n,p_x}$ and $\Phi_{\alpha n,p_x-q}$ have the relative shift $\Delta y=dB_y/B_z$ in real space, as shown in Eq.~(\ref{Dy}).  Since the wave functions in Eq.~(\ref{eigenfunc1}) oscillate in real space on the scale of $l/\sqrt{n}$, the overlap between $\Phi_{n,p_x}$ and $\Phi_{\alpha n,p_x-q}$ oscillates as a function of $B_y$, resulting in the oscillatory SAS splitting of the Landau levels in Figs.~\ref{fig:same}(a) and \ref{fig:opposite}(a).  For a sufficiently strong $B_y$, the distance $\Delta y$ exceeds the width $l\sqrt n$ of the Landau wave functions, so the overlap matrix elements $w_{n,\alpha n}$ vanish, and the Landau levels (\ref{sas}) become degenerate. The positions of the nodes, where the SAS splitting vanishes, are different in Figs.~\ref{fig:same}(a) and \ref{fig:opposite}(a) for $\alpha=\pm1$ reflecting the difference between $w_{n,n}$ and $w_{n,-n}$.  In Sec.~\ref{sec:Semiclassical}, we show that it is a consequence of different Berry phase contributions.

The lines in Figs.~\ref{fig:same}(a) and \ref{fig:opposite}(a) separate regions where the Hall conductivity has the quantized values $\sigma_{xy}=\nu e^2/h$ indicated on the plots, assuming that all Landau levels are filled below the energy $E$.  For two decoupled Dirac layers in the spinless case, the filling factor runs through the odd integers $\nu = 2j+1$, where $j$ is integer.  However, in the presence of the coupling $w$ between the Dirac layers, the even filling factors $\nu=2j$ becomes available for the energies inside the SAS splitting, which oscillates as a function of $B_y$.

In Figs.~\ref{fig:same}(b) and \ref{fig:opposite}(b), we plot the same data in a different way.  We fix the chemical potential, so that the Fermi energy $E_F=vp_F$ and the Fermi momentum $p_F$ are constant, and plot a map of the density of states (DOS) at the Fermi level as a function of $B_y$ and $B_z$.  Figures~\ref{fig:same}(b) and \ref{fig:opposite}(b) exhibit peaks in DOS when the Landau levels cross the chemical potential.  The Landau level index $n$ is indicated on the right vertical axis.  For $B_y=0$ and increasing $B_z$, the Landau levels with the indices $n\propto 1/B_z$ cross the Fermi energy.  For increasing $B_y$, the SAS splitting between the Landau levels oscillates and passes through a series of nodes.  A similar oscillatory SAS splitting was observed experimentally in semiconducting bilayers with a parabolic dispersion relation \cite{Gusev2008}.  In the regions between the peaks in DOS, the Hall conductivity has the quantized values $\sigma_{xy}=\nu e^2/h$ indicated in Figs.~\ref{fig:same}(b) and \ref{fig:opposite}(b).  Thus, in the double layer geometry, transitions between the quantum Hall plateaus can be driven by both the in-plane and out-of-plane components of the magnetic field.

For low magnetic fields, the SAS splitting nodes align along the dashed lines corresponding to the ``magic'' tilt angles $\theta_N$ .  In order to find these angles, let us examine where the diagonal tunneling matrix elements
\begin{equation}
  w_{n,\alpha n} = \frac{w e^{-\beta^2/2}}{2}
  \left[ L^{(0)}_{|n|}\left(\beta^2\right)
  +\alpha L^{(0)}_{|n|-1}\left(\beta^2\right) \right]
\label{tun}
\end{equation}
vanish. Using the asymptotic approximation of the Laguerre polynomials for $n\gg x \gg1$
\begin{equation}
  L_n^{(k)}(x) \approx
  \frac{n^{\frac k2-\frac14}e^{\frac x2}}{\sqrt\pi x^{\frac k2+\frac14}}
  \cos\left[2\sqrt{nx}-\frac{\pi}{2}\left(k+\frac12\right)\right]
\label{n>>1}
\end{equation}
in Eq.~(\ref{tun}) for $\alpha=1$, we find
\begin{align}
  \frac{w_{n,n}}{w} & =
  \frac{\cos\left(2\sqrt n\beta-\frac\pi4\right)}{\sqrt{\sqrt{n}\beta\pi}}
  = \frac{\cos\left(\frac{p_F d\tan\theta}{\hbar}-\frac\pi4\right)}
  {\sqrt{\pi p_Fd\tan\theta/2\hbar}}.
\label{t1}
\end{align}	
Here we kept only the leading terms in $1/n$.  Assuming that the Landau level $n$ is at the chemical potential $E_n=vp_F$, we expressed the Fermi momentum as $p_F=\sqrt{2ne\hbar B_z}$, so that
\begin{equation}
  2\sqrt n\beta = \frac{p_F d\tan\theta}{\hbar}, \qquad
  \frac{\beta}{2\sqrt n} = \frac{eB_y d}{2p_F}.
\end{equation}
For the opposite chiralities $\alpha=-1$, using the identity $L^{(0)}_{n+1}(x)-L^{(0)}_{n}(x)=-\frac{x}{n+1}L^{(1)}_n(x)$ and the asymptotic formula (\ref{n>>1}), we obtain
\begin{align}
  \frac{w_{n,-n}}{w} & = \sqrt{\frac{\beta}{4n^{3/2}\pi}}
  \cos\left( 2\sqrt n \,\beta-\frac{3\pi}{4}\right)
\nonumber \\
  & = \sqrt{\frac{e^2\hbar B_yB_z d}{2\pi p_F^3}}
  \,\cos \left( \frac{p_F d\tan\theta}{\hbar} - \frac{3\pi}{4} \right).
\label{t2}
\end{align}
The arguments of the cosine functions in Eqs.~(\ref{t1}) and (\ref{t2}) are different, so the matrix elements $w_{n,n}$ and $w_{n,-n}$ vanish at the different magic tilt angles $\theta_N$
\begin{eqnarray}
  & p_F d\tan\theta_N = \hbar\left(\pi N - \frac\pi4 \right),
  \quad &\alpha=+1,
\label{zeros1} \\
  & p_F d\tan\theta_N = \hbar\left(\pi N + \frac\pi4\right), \quad & \alpha=-1.
\label{zeros2}
\end{eqnarray}
Equation~(\ref{zeros1}) is equivalent to Eq.~(\ref{classic}) for the parabolic dispersion. Note that the condition~(\ref{zeros2}) was also obtained in Ref.~[\onlinecite{Yagi1994}] for a three-dimensional material with an azimuthally corrugated Fermi surface.  Our result~(\ref{zeros2}) does not depend on the azimuthal direction of the in-plane magnetic field and, thus, can be experimentally distinguished from the scenario proposed in Ref.~[\onlinecite{Yagi1994}]. The magic angles $\theta_N$ given by Eqs.~(\ref{zeros1}) and (\ref{zeros2}) are shown by the dashed lines in Figs.~\ref{fig:same} and \ref{fig:opposite}, correspondingly.  We observe that the SAS splitting nodes align very well with these lines for moderate magnetic fields.  For stronger fields, the magic angles become dependent on the magnitude of the field.

Angular dependence of the Landau levels can be also observed in the out-of-plane conductance $G_{zz}=dI_z/dV_z$ in a tilted magnetic field. In the tunneling formalism for small $w$, the tunneling conductance is proportional to
\begin{equation}
  G_{zz} \propto B_z|w_{n,\alpha n}|^2\,\rho^2_n(E_F),
\label{tunnel}
\end{equation}
where $\rho_n(E_F)$ is the DOS for the original unperturbed Landau level  (\ref{eigenfunc1}) at the Fermi energy, as discussed in Appendix~\ref{Sec-TunnelConduct}.  The tunneling conductance $G_{zz}$ is plotted in panels (c) of Figs.~\ref{fig:same} and \ref{fig:opposite} as a function of both $B_y$ and $B_z$.  Comparing panels (b) and (c), we observe that $G_{zz}$ has maxima where the SAS splitting is large.  Conversely, the tunneling conductance is suppressed at the magic angles defined by Eqs.~(\ref{zeros1}) and (\ref{zeros2}) and shown by the dashed lines.  The oscillations of $G_{zz}$ as a function of $B_z$ for a fixed $B_y$ represent the usual Shubnikov-de Haas oscillations, whereas the oscillations of $G_{zz}$ as a function of the tilt angle $\tan\theta=B_y/B_z$ represent AMRO.

As indicated above Eq.~(\ref{n>>1}), the approximation for the Laguerre polynomials is applicable only for the high Landau levels with $n\gg1$, i.e.,\ for weak magnetic fields $B_z$.  Moreover, it is also required that $n\gg\beta^2$, which means a weak magnetic field $B_y$ such that $eB_yd\ll2p_F$ \footnote{In 2D Dirac materials, the Fermi energy and Fermi momentum can be tuned by external gating.  Using the graphene Fermi velocity $v=10^6\,{\rm m/s}$ and the interlayer distance $d=2\,{\rm nm}$, we estimate the magnitude of the magnetic field $B_y = 2p_F/ed=2E_F/evd$ where the Fermi circles detach as $B_y=10$~T and $100$~T for $E_F=10$~meV and $100$~meV.}.

For stronger magnetic fields, Eq.~(\ref{tun}) should be used without approximations.  On the horizontal axes in panels (b) and (c) in Figs.~\ref{fig:same} and \ref{fig:opposite}, the value $eB_yd/p_F=2$ corresponds to detachment of the Fermi circles in the two layers displaced by $q$, as shown at the top of panels (c).  For $\alpha=1$, the effective interlayer coupling, as measured by the SAS splitting and tunneling conductance $G_{zz}$, is maximal for $B_y=0$ and is suppressed around $eB_yd\approx2p_F$.  This is because the spinor wave functions (\ref{eigenfunc1}) are orthogonal at the opposite sides of the Fermi circle.  In contrast, for $\alpha=-1$, the effective interlayer coupling is suppressed around $B_y\approx0$ and is maximal for $eB_yd\approx2p_F$, because the spinors (\ref{eigenfunc1}) have opposite chiralities in this case.  Panels (b) and (c) in Fig.~\ref{fig:opposite} demonstrate an interesting pattern of magnetic oscillations versus $B_y$ and $B_z$ around $eB_yd\approx2p_F$.  This pattern originates from quantization of the electron orbits around the unshaded area ACFG in Fig.~\ref{fig:Bilayer}(b), which shrinks when $eB_yd\to2p_F$.  A similar pattern of magnetic oscillations versus $B_y$ and $B_z$ was observed experimentally \cite{Harff1997} in semiconducting bilayers with population imbalance between the layers.

The first-order perturbation theory in Eq.~(\ref{sas}) is applicable when the SAS splitting $w_{n,\alpha n}$ is smaller than the energy difference between consecutive  Landau levels.  Otherwise, the full equation (\ref{Schrod}) with the off-diagonal matrix elements $w_{n,m}$ should be solved numerically.  However, it is also possible to get an insight using the semiclassical approximation described below.

\section{Semiclassical description} \label{sec:Semiclassical}

\begin{table*}
  \begin{tabular*}{.95\textwidth}{@{\extracolsep{\fill}} c  c  c  c  c }
  \hline\hline
  Type of Dirac Hamiltonian                      & $\alpha = 1$, $W=w\bm I$    & $\alpha = -1$, $W = w\bm I$    &   $\alpha = 1$, $W =w(\sigma_x+i\sigma_y)$               &  $\alpha = 1$, $W = w \sigma_x $ \\ 
  Physical system                                & Double layer graphene         & TI film                        &   Bernal-stacked graphene                              &                                           \\ \hline
  $\Gamma^2(ADF)-\Gamma^1(ACF)$                  & $\chi$                        & $\pi$                          &   $\chi$                                               &  $\chi$                                   \\  
  ${\rm Arg}\,W^{21} (A)-{\rm Arg}\,W^{21}(F)$   & $2\pi-\chi$                   & $0$                            &   $\pi-\chi$                                           &  $0$                              \\  
  $\varphi\,{\rm mod}\,2\pi$                       & $0$                           & $\pi$                        &   $\pi$                                                &  $\chi$                                   \\ \hline\hline
\end{tabular*}
\caption{The phase shift $\varphi$ given by Eq.~(\ref{phase}), which appears in Eqs.~(\ref{interf1}) and (\ref{interf2}), for different types of Dirac Hamiltonians in the top row and the corresponding physical systems in the second row.  The variable $\alpha$ represents relative chirality of the Dirac cones in Eq.~(\ref{Ham}), whereas $W$ is the tunneling matrix in Eq.~(\ref{w}).   The angle $\chi$ is shown in Figs.~\ref{fig:Bilayer}(b) and \ref{fig:Cones}.  The total phase $\varphi$ in the last row is the sum of the third and fourth rows representing the Berry-phase (\ref{gam}) and the tunneling (\ref{w21}) contributions to Eq.~(\ref{phase}). }
\label{Table}
\end{table*}

Here we discuss how to derive the magic angles in the semiclassical approximation. Let us first review the semiclassical arguments in the case where the layers have a parabolic in-plane spectrum \cite{Cooper2006,Yakovenko2006}. As illustrated in Fig.~\ref{fig:Bilayer}(b) (as well as in Fig.~\ref{fig:Cones}), the in-plane magnetic field $B_y$ shifts the Fermi momenta by $q$, whereas the perpendicular magnetic field $B_z$ induces cyclotron motion in momentum space. Then, interference between the paths $ADF$ and $ACF$ determines the effective coupling between the layers. Similarly to the semiclassical Onsager quantization \cite{Mikitik1999,Carmier2008,Fuchs2010}, the interference is controlled by the shaded area $S_p$ between the two paths in Fig.~\ref{fig:Bilayer}(b)
\begin{equation}
  \frac{S_p}{e\hbar B_z} + \varphi = \frac{\pi}{2} -\pi + 2\pi N.
\label{interf1}
\end{equation}
Here, the term $\pi/2$ originates from the Maslov index at the turning points, whereas the term $-\pi$ represents destructive interference. For a small shift $q\ll p_F$, the area becomes $S_p=2p_Fq=2p_FeB_yd$, so the destructive interference condition (\ref{interf1}) becomes 
\begin{equation}
	p_F d \tan\theta_N =\hbar\left(\pi N-\frac{\pi}{4}-\frac{\varphi}{2}\right),\quad N = 1,2,\ldots
\label{interf2}
\end{equation}
For the in-plane parabolic energy dispersion $h(\bm p)=p^2/2m$, the extra phase $\varphi$ vanishes, i.e. $\varphi=0$, and Eq.~(\ref{interf2}) reproduces Eq.~(\ref{classic}).

For the in-plane Dirac Hamiltonian (\ref{h}), the spinor eigenstates~(\ref{wv}) produce an additional phase \cite{Mikitik1999,Carmier2008,Fuchs2010}
\begin{align}
	\varphi &= \Gamma^2(ADF)-\Gamma^1(ACF) +{\rm Arg}\,W^{21} (A)-{\rm Arg}\,W^{21}(F), \label{phase}
\end{align}
where the upper indices $j=1,2$ denote the layer number. The first two terms represent the Berry phases
\begin{align}
	\Gamma^j(\mathcal C) = i\int\limits_{\mathcal C} d\bm p\,\langle \psi^j_{\bm p} \mid \partial_{\bm p} \mid \psi^j_{\bm p}\rangle \label{gam}
\end{align}
accumulated during the semiclassical motion along the paths $ADF$ or $ACF$, denoted by the symbol $\mathcal C$ for brevity.  The last two terms in Eq.~(\ref{phase}) describe the phases picked during the inter-orbit tunneling
\begin{equation}
	W^{21}(X) = \langle \psi^2_{\bm p_X}\mid W \mid \psi^1_{\bm p_X} \rangle,
	\label{w21}
\end{equation}
where $X$ denotes the intersection points $A$ and $F$ for brevity.  In contrast to Eq.~(\ref{w}), we now allow for an arbitrary interlayer tunneling matrix $W$.  Note that the phase~$\varphi$ does not depend on a particular choice of the gauge for the eigenstates~(\ref{wv}), although the individual terms in Eq.~(\ref{phase}) are gauge-dependent.

Let us calculate the phase $\varphi$ for the case, where $\alpha = 1$ and $W=w\bm I$ considered in the previous section. For the wave functions~(\ref{wv}), the Berry phase~(\ref{berry}) is given by the half of the arc traced by the orbit as viewed from the origin. So, we obtain the Berry phase contribution $\Gamma^2(ADF)-\Gamma^1(ACF)=\chi$ expressed via the angle $\chi$ shown in Figs.~\ref{fig:Bilayer}(b) and \ref{fig:Cones}. On the other hand, the contribution of tunneling in Eq.~(\ref{phase}) is ${\rm Arg}\,W^{21} (A)-{\rm Arg}\,W^{21}(F)=2\pi-\chi$. We sum the Berry phase and tunneling contributions and obtain $\phi = 2\pi$. Thus, the interference condition~(\ref{interf2}) recovers Eq.~(\ref{zeros1}). For the case of $\alpha=-1$ and $W=w\bm I$, which corresponds to a TI film, the Berry phase contribution is $\Gamma^2(ADF)-\Gamma^1(ACF)=\pi$, whereas the tunneling contribution vanishes ${\rm Arg}\,W^{21} (A)-{\rm Arg}\,W^{21}(F)=0$. Thus, we substitute $\varphi=\pi$ in Eq.~(\ref{interf2}) and reproduce Eq.~(\ref{zeros2}).  In the Bernal-stacked graphene bilayer,
the interlayer tunneling matrix $W=w(\sigma_x+i\sigma_y)$ couples one sublattice of one layer to another sublattice of another layer\cite{Pershoguba2010} for the Dirac cones of the same chirality (so $\alpha=1$).  In this case, we also obtain the phase $\varphi=\pi$.  For a hypothetical tunneling matrix $W=w \sigma_x$, we obtain the phase $\varphi=\chi=2\,{\rm arcsin}(q/2p_F)$, which depends on the in-plane magnetic field via $q=eB_yd$.  These results are summarized in Table~\ref{Table}.  The phase $\varphi$ strongly depends on the interlayer tunneling matrix~$W$ and the relative chirality $\alpha$ of the coupled Dirac cones.

The above discussion is applicable when the out-of-plane magnetic field $B_z\gg B_0$ is stronger than the magnetic breakdown field $B_0$. In general, the  interlayer tunneling amplitude $w$ hybridizes and splits the electron orbits at the intersection points A and F in Fig.~\ref{fig:Bilayer}(b).  Below the magnetic breakdown field at $B_z\ll B_0$, the electrons predominantly move along the hybridized orbits ACFG and ADFH, called the ``lens'' and ``peanut'' in Ref.~[\onlinecite{Harff1997}], and have a small probability $P=\exp(-B_0/B_z)$ of changing the orbit.  In the opposite limit $B_z\gg B_0$ above the magnetic breakdown, the electrons predominantly stay on the circular orbits within each layer and have a small probability $P=1-\exp(-B_0/B_z)\approx B_0/B_z$ of tunneling to another layer at the intersection points A and F in Fig.~\ref{fig:Bilayer}(b). The magnetic breakdown field \cite{MacDonald1992} is given by the following expression
\begin{align}
  B_0 = \frac{2\pi {w'}^2}{\hbar ev^2\sin \chi} =
  \left\{\begin{array}{ll}
  \frac{2\pi p_F w^2}{\hbar e v^2 q}\sqrt{1-\frac{q^2}{4p_F^2}}, & \alpha=+1, \\
  & \\
  \frac{\pi q w^2}{2\hbar e v^2 p_F\sqrt{1-\frac{q^2}{4p_F^2}}}, & \alpha=-1,
  \end{array}\right.
\label{breakdown}
\end{align}
as discussed in Appendix~\ref{Sec-breakdown}. Here $\chi$ is the intersection angle of the two cyclotron orbits in Fig.~\ref{fig:Cones}, and $w'$ is the effective coupling between the orbits. The angle $\chi$ can be expressed via the in-plane magnetic shift $q$ as $\sin(\chi/2)=q/2p_F$, and $w'$ is determined by the spinor structure of the wave functions in Eq.~(\ref{wv}).  For $\alpha =1$, the angle between the pseudospins on different orbits is $\chi$, so the effective coupling is $w'=w\cos(\chi/2)$. For $\alpha=-1$, the angle between the pseudospins is $\pi-\chi$, so the effective coupling is $w'=w\sin(\chi/2)$.

\section{Experimental relevance and conclusions}

Among the Dirac materials, AMRO have been observed experimentally in the intercalated graphite \cite{Iye1994} at the angles close to $\theta=\pi/2$ where the magnetic field is almost parallel to the layers.  This is because $\tan\theta_N\propto1/p_Fd$ is large for a small interlayer distance $d$ and a small Fermi momentum $p_F$.  In the graphene double layer reported in Ref.~[\onlinecite{Mishchenko2014}], the interlayer distance $d=1.4$~nm includes the boron nitride layers separating the two graphene layers. Taking the Fermi energy as $E_F=0.2$~eV and using the Fermi velocity $v=10^6\,{\rm m}/{\rm s}$, we find the Fermi momentum $p_F/\hbar=E_F/\hbar v=3\times10^8{\rm\,m}^{-1}$.  Using Eq.~(\ref{zeros1}), we estimate the first magic angle as $\theta_1 = \arctan(3\pi\hbar/4p_Fd) = 80^{\circ}$. Taking the interlayer coupling to be $w\sim10$ meV and $\sin\chi\sim1$ in Eq.~(\ref{breakdown}), we estimate the magnetic breakdown field as ${B_0\sim1\,{\rm T}}$. Thus, we conclude that observation of AMRO in the graphene double layer of Ref.~[\onlinecite{Mishchenko2014}] in a tilted magnetic field is experimentally feasible.

In conclusion, in this paper we examined the effects of a tilted magnetic field in the Dirac double layer.  We derived the general equation (\ref{Schrod}) for the electron energy spectrum and its approximations (\ref{sas}) and (\ref{tun}) for a sufficiently small interlayer tunneling amplitude $w$.  We found that the SAS energy splitting between the Landau levels oscillates as a function of the in-plane magnetic field $B_y$ and vanishes at the series of ``magic'' tilt angles $\theta_N$ of the magnetic field given by Eqs.~(\ref{zeros1}) and (\ref{zeros2}).  The interlayer tunneling conductance (\ref{tunnel}) is suppressed at these magic angles.  Our results generalize the previously known phenomenon of the angular magnetoresistance oscillations (AMRO) to the Dirac double layers, where the magic angles depend on the Berry phases and coupling between the Dirac cones: see Eqs.~(\ref{interf1})-(\ref{phase}). Our theoretical results are applicable to, e.g., graphene double layers and thin films of topological insulators studied experimentally in Refs.~[\onlinecite{Mishchenko2014}] and [\onlinecite{Armitage}], respectively. We also found that the quantum Hall conductivity $\sigma_{xy}$ depends on both $B_y$ and $B_z$ components of the magnetic field, as indicated by the blue and white numbers in the panels (a) and (b) of Figs.~\ref{fig:same} and \ref{fig:opposite}. It would be interesting to further explore the role of interactions in the quantum Hall regime in the tilted field geometry \cite{Zhao2010}.

{\it Acknowledgment}. T
his work was supported by ERC DM-321031 and US DOE BES E304 (S.S.P., D.S.L.A., and A.V.B). We would like to thank Tim Khodkov and Yaron Kedem for helpful discussions.

\appendix

\section{CALCULATION OF MATRIX ELEMENTS} \label{Sec-MatrixElements}

Here we calculate the matrix element in Eq.~(\ref{wnm}). Using the spinor structure of the wave functions~(\ref{eigenfunc1}) and assuming that $|n|\ge|m|$, we write
\begin{align}
  & w_{n,m}/w = \,\langle\Phi_{n,p_x}  \mid\Phi_{m,p_x-q}\rangle
\label{me1} \\
  & = \left\{ \begin{array}{ll}
  (M_{|n||m|}+{\rm sgn}(nm)M_{|n|-1\,|m|-1})/2, & |m|>0, \\
  M_{|n|0}/\sqrt{2}, & |n|>m=0,  \\  M_{00}, & n=m=0,
  \end{array} \right. \nonumber	
\end{align}
where
\begin{equation}
	M_{|n||m|}=\langle\phi_{|n|,p_x}\mid\phi_{|m|,p_x-q}\rangle \label{me2}
\end{equation}
is the matrix element between the shifted harmonic-oscillator functions.  As discussed in Sec.~\ref{sec:magfields}, the shift in momentum $\Delta p_x=-q$ corresponds to the spatial shift by $\Delta y=q/eB_z=ql^2/\hbar$. So, the matrix element (\ref{me2}) can be expressed via the translation operator $\hat p_y=-i\hbar\partial_y$:
\begin{align}
  M_{|n||m|}=\langle\phi_{|n|,p_x}\mid e^{i \hat p_y q l^2/\hbar^2}
  \mid\phi_{|m|,p_x}\rangle,
\label{eqS1}
\end{align}
where $\hat p_y=\hbar(\hat a-\hat a^\dag)/i\,l\sqrt 2$ is written in terms of the lowering and raising operators.  Then we use the Baker-Hausdorff formula to decouple the operators in the exponent
\begin{eqnarray*}
  M_{|n||m|} &&=\langle\phi_{|n|,p_x}\mid e^{(\hat a-\hat a^\dag)\beta}
  \mid\phi_{|m|,p_x} \rangle \\
  &&=e^{-\beta^2/2}\langle\phi_{|n|,p_x}\mid e^{-\hat a^\dag \beta} e^{\hat a\beta}
  \mid\phi_{|m|,p_x} \rangle,
\end{eqnarray*}
where the parameter $\beta$ is defined in Eq.~(\ref{beta}). Expanding the exponential functions and using the algebra of the raising and lowering operators, we obtain
\begin{align}
	M_{|n||m|} & = e^{-\beta^2/2} (-\beta)^{|n|-|m|} \\
	& \times \sqrt{\frac{|n|!}{|m|!}} \sum_{k=0}^{|m|} \frac{(-\beta^2)^k\, |m|\ldots (|m|-k+1)}{k!(|n|-|m|+k)!} \nonumber \\
	& = e^{-\beta^2/2}(-\beta)^{|n|-|m|} \sqrt{\frac{|m|!}{|n|!}} L^{(|n|-|m|)}_{|m|} \left(\beta^2\right), \nonumber
\end{align}
where we use the definition of the Laguerre polynomials in the last line.

The matrix elements for $|m|>|n|$ are obtained by interchanging $n$ and $m$ and altering the sign $\beta\rightarrow-\beta$.

\section{DERIVATION OF TUNNELING CONDUCTANCE} \label{Sec-TunnelConduct}

Here we give a brief derivation of the out-of-plane tunneling conductance~(\ref{tunnel}) between the two layers. In the tunneling-current formalism \cite{BruusFlensberg} for  small interlayer coupling $w$, we write
\begin{equation}
	G_{zz} =\frac{dI_z}{dV_z} =  \frac{2\pi e^2}{\hbar} \sum_{n,m,p_x}
  |w_{n,\alpha m}|^2\, \rho_n(E_F)\,\rho_{\alpha m}(E_F),
\end{equation}
where $n,m$ are the integers labeling the Landau wave functions on the different layers, and $w_{n,m}$ are the tunneling matrix elements (\ref{wnm}). In the chosen gauge, the momentum $p_x$ defines the coordinate $y=-p_x/eB_z$ around which the Landau wave functions are localized, as discussed in Section~\ref{sec:magfields}. Thus, for a double layer of the finite size $L_x$ and $L_y$, we have
\begin{equation}
  \sum_{p_x} \rightarrow \frac{L_x}{2\pi\hbar}
  \int\limits_{-eB_zL_y/2}^{eB_zL_y/2} dp_x = \frac{eB_zL_xL_y}{2\pi\hbar},
\end{equation}
where $L_x$ defines the normalization of the differential $dp_x$, whereas $L_y$ defines the limits of integrations. So, the tunneling conductance becomes
\begin{equation}
	G_{zz} = \frac{e^2}{\hbar} \frac{eB_z L_x L_y}{\hbar} \sum_{n,m} |w_{n,\alpha m}|^2\,
  \rho_n(E_F)\,\rho_m(E_F),
\label{G1}
\end{equation}
Note, that the second fraction containing the magnetic field $B_z$ represents the degeneracy of the Landau levels. We assume that DOS of the Landau level $n$ has a finite width $\Gamma$\cite{Grigoriev2014}
\begin{equation}
  \rho_n(E) = \frac{1}{\sqrt \pi \Gamma}\exp\left[ -\frac{(E-E_n)^2}{\Gamma^2} \right].
\label{lldos}
\end{equation}
If the width $\Gamma\ll|E_n-E_{n\pm1}|$ is much smaller than the energy difference between consecutive Landau levels, the tunneling conductance~(\ref{G1}) can be approximated as
\begin{equation}
  G_{zz} = \frac{e^2}{\hbar} \frac{eB_z L_x L_y}{\hbar}  |w_{n,\alpha n}|^2 \rho^2_n(E_F),
\end{equation}
thus producing Eq.~(\ref{tunnel}).

The effect of the Landau levels DOS profile on AMRO was studied in Ref.~[\onlinecite{Grigoriev2014}].  Reference~[\onlinecite{Grigoriev2014}] also contains a derivation of the tunneling conductance for a large Landau level broadening $\Gamma\gg|E_n-E_{n\pm1}|$.

\section{DERIVATION OF THE MAGNETIC BREAKDOWN FIELD} \label{Sec-breakdown}

Here we derive Eq.~(\ref{breakdown}) for the magnetic breakdown field using the Landau-Zener theory of tunneling.  The Fermi circles corresponding to different layers intersect at the angle $\chi$ at the points A and F in Figs.~\ref{fig:Bilayer}(b) and \ref{fig:Cones}. In the vicinity of, e.g., point A in the momentum space, the effective Hamiltonian of the double layer in the basis $(\psi^1,\psi^2)$ can be approximated as
\begin{equation}
  \left[\begin{array}{cc} (\bm p-\bm p_{\rm A})\cdot\bm v_1 & w' \\
  w' & (\bm p-\bm p_{\rm A})\cdot\bm v_2
  \end{array}\right],	
\end{equation}
where $\bm v_1$ and $\bm v_2$ are the local velocities of the two orbits at the point A, and $w'$ is the local effective coupling. It is convenient to use the reference frame in momentum space where $\bm p_{\rm A}=0$, and the $x$ axis bisects the angle $\chi$. Then the velocities are $\bm v_1 =v\left(-\sin\frac{\chi}{2},\cos\frac{\chi}{2}\right)$ and  $\bm v_2 =v\left(\sin\frac{\chi}{2},\cos\frac{\chi}{2}\right)$, and the Hamiltonian becomes
\begin{equation}
  \left[\begin{array}{cc} -vp_x\sin\frac{\chi}{2}+vp_y\cos\frac{\chi}{2}
  & w' \\ w' & vp_x\sin\frac{\chi}{2}+vp_y\cos\frac{\chi}{2}
  \end{array}\right].
\label{localHam}	
\end{equation}
In the perpendicular magnetic field $B_z$ described by the gauge $\bm A = -yB_z\hat{\bm x}$, the momenta become $(p_x,p_y)\rightarrow(p_x +eB_zy,p_y)$.  Quasiclassical dynamics of a wave packet moving in the top layer is governed by the upper-diagonal element of Hamiltonian~(\ref{localHam})
\begin{equation}
	h_1 = -v(p_x +eB_zy)\sin\frac{\chi}{2}+vp_y\cos\frac{\chi}{2}.
\end{equation}
The classical equations of motion can be integrated
\begin{equation*}
  \begin{array}{ll}
  \dot y = \frac{\partial h_1}{\partial p_y} =  v\cos\frac{\chi}{2},
     & \Rightarrow \;\;  y(t) = t\,v\cos\frac{\chi}{2}-\frac{p_x}{eB_z}, \\
  \dot p_y =-\frac{\partial h_1}{\partial y}= evB_z\sin\frac{\chi}{2},
  & \Rightarrow \;\;  p_y(t) = t\,evB_z\sin\frac{\chi}{2},
  \end{array}
\end{equation*}
where the initial conditions are chosen so that $h_1(0)=0$. Substituting these solutions into the double-layer Hamiltonian~(\ref{localHam}), we find the Landau-Zener Hamiltonian with the time-dependent lower diagonal element
\begin{equation}
  \left[\begin{array}{cc} 0 & w' \\ w' & t\,ev^2B_z\sin\chi
  \end{array}\right].
\label{localHam1}	
\end{equation}
According to the Landau-Zener formula, the probability that the wave packet stays on the same orbit $\psi_1$ is
\begin{equation}
  P = \exp\left( -\frac{2\pi w'^2}{\hbar ev^2B_z\sin\chi} \right)
  = \exp\left( -\frac{B_0}{B_z} \right),
\label{probab}
\end{equation}
where $B_0$ is the magnetic breakdown field
\begin{equation}
	B_0 = \frac{2\pi w'^2}{\hbar ev^2\sin\chi}.
\end{equation}

The above consideration is applicable to double layers with both parabolic and Dirac energy dispersion.  However, in the Dirac case, the effective tunneling $w'$ is determined by the scalar product of the spinor wave functions~(\ref{wv}) in the opposite layers.  The angle between the pseudospins is $\chi$ for $\alpha=1$ and $\pi-\chi$ for $\alpha=-1$, so the effective couplings are $w'= w\cos(\chi/2)$ and $w'=w\sin(\chi/2)$, respectively.  We further express the angle $\sin(\chi/2)=q/2p_F$ via the magnetic shift $q$ and obtain Eq.~(\ref{breakdown})
\begin{align}
  B_0 =
  \left\{\begin{array}{ll}
  \frac{2\pi p_F w^2}{\hbar e v^2 q}\sqrt{1-\frac{q^2}{4p_F^2}}, & \alpha=+1, \\
  & \\
  \frac{\pi q w^2}{2\hbar e v^2 p_F\sqrt{1-\frac{q^2}{4p_F^2}}}, & \alpha=-1.
  \end{array}\right.
\end{align}
\vspace*{5cm}
\bibliographystyle{apsrev4-1}
\bibliography{DiracAMRO}

\end{document}